\documentclass[12pt]{article}
\usepackage{multirow}
\usepackage{amsmath,amsfonts,amssymb,bm,slashed,bbm,mathrsfs}
\usepackage{graphicx,color}
\renewcommand{\Ref}[1]{(\ref{#1})}
\newcommand{\eq}[2]{\begin{align}\label{#1}#2\end{align}}

\newcommand{\nn}{\nonumber}
\renewcommand{\ni}{\noindent}
\newcommand{\pa}{\partial}

\newcommand{\sig}{\sigma}

\begin{document}

\title{Spontaneous  magnetization of a vacuum in high temperature gluodynamics (two-loop approximation) }
\author{
 V. Skalozub\thanks{e-mail: Skalozub@ffeks.dnu.edu.ua}\\
{\small Oles Honchar Dnipro National University, 49010 Dnipro, Ukraine}}

\date{ }
\maketitle\thispagestyle{empty}
%
\begin{abstract}
In SU(N) gluodynamics, at high temperature the spontaneous  magnetization, $b(T) \not = 0$, of a vacuum happens in the approximation to the effective potential - the tree plus the one-loop, plus daisy diagrams, $W(b)= \frac{b^2}{2 g^2} + W^{(1)}(b) + W^{daisy}(b) $. At the same time, in two-loop approximation,  $ W(A_0)= W^{(1)}(A_0) + W^{(2)}(A_0), $ other classical field - $A_0$ condensate  directly related to the Polyakov loop - is also spontaneously generated. To investigate the creation of the condensates together,   the two loop effective potential of both fields should be calculated. This program was realized recently for SU(2) in \cite{bord22-82-390}.  However, the generation of magnetic field in two-loop order was not studied in detail. In the present paper, we  compute the value of chromomagnetic field $b(T)$ for latter case. Then, considering the spectrum of color charged gluons at the  background of both condensates, we conclude that the $A_0$ stabilizes the magnetized vacuum at high temperature. 	This is in agreement with the lattice simulations carried out already and clarifies the mechanism of the magnetic field stabilization.

\end{abstract}

\section{\label{T1}Introduction}
Deconfinement phase transition (DPT)  as well as the properties of quark-gluon plasma (QGP) are widely investigated  for many years. Most results have been obtained in the lattice simulations because of a large coupling value $g \ge 1$ at the  phase transition temperature $T_c$. But at high temperatures due to asymptotic freedom the analytic methods are also reliable. They give a possibility for investigation various phenomena in the plasma. Among them is the creation of gauge field condensates described by the classical solutions to field equations without sources. Only such type fields could appears spontaneously. The well known ones are the so-called $A_0$ condensate, which is algebraically related to the Polyakov loop, and chromomagnetic field $b(T)$, which is the Savvidy vacuum state at high temperature. These condensates result in numerous effects  and could serve as the signals of the DPT. The  $A_0$ condensation has been investigated by  different methods. For a recent work see \cite{gaof21-103-094013} and references  therein.

Although these condensates are the consequences of asymptotic freedom, they are generated at different orders  in coupling constant or the number of loops for the effective potential. This is the reason why they have different temperature dependencies and could play different  roles.   For example, $A_0$ is generated in $g^4$ order of coupling constant and determined by the ratio of two- and one-loop  contributions to $W(A_0)$. So it has  the $g^2$ order. The field $b(T)$ is generated in one-loop plus daisy approximation and also has the order $g^2$ in coupling constant. But it has other temperature dependence due to a tree-level contribution of classical field equations. This   is important at high temperature. All mentioned features  require special comprehensive  considerations.

The fields investigated below are an important topic towards a theory of confinement. The $A_0$-background is relevant because at finite temperature such field cannot be gauged away and is intensively investigated beginning with \cite{weis82-25-2667}. In the early 90-ies, two-loop contributions were calculated in QCD and with these,  the effective potential has non-trivial minimums and related condensate fields (see, for instance, \cite{skal94-57-324}).  They form a hexagonal structure in the plane of the relevant color components $A_0^3$ and $A_0^8$ of the background field.

A different kind of background is the chromomagnetic one. More details about this field and the ways of its stabilization at finite temperature can be found, in particular,  in  \cite{skal00-576-430}. The magnetization  is also resulted from the minimum of the effective potential which is stable in the approximation of one loop plus daisies. A common generation of both fields was studied in \cite{bord22-82-390}. Here, new integral representation, which generalized the known integral representation for the Bernoulli polynomials, was introduced and admitted introducing either $A_0$ or $b$ fields up to two loop order. Within this representation, in particular, the known results for separate generation of the fields have been reproduced. However, the spontaneous generation of chromomagnetic field up to two-loop order was not investigated in detail. Therefore, the mechanism of the vacuum stabilization remained not clarified. This is the problem which is investigated in the present paper.
This point  is of grate importance because in the lattice calculations
  accounting for both backgrounds  \cite{demc13-21-13} it was observed that in the presence of a constant color magnetic field the Polyakov loop acquires  a non-trivial spatial structure    along the   direction of the field. More interesting, in  \cite{demc08-41-165051}  a common spontaneous generation of both fields was detected.

As a step towards the simultaneous generation of both background fields in a perturbation  approach, we consider both these  fields on the two-loop level. More specifically, we calculate the effective potential as a function of both parameters, $A_0$, and $b$, in $SU(2)$ gluodynamics. The integral expressions for the effective potential of the $A_0$  are generalized  to include the magnetic background in two-loop order.
Also, we consider the limiting cases $A_0=0$ and $b \not =0$ and find, for instance, the magnetic condensate in two-loop order, which was also considered in  \cite{bord22-82-390},  \cite{skal00-576-430} but using other approaches and in not wide temperature interval.
Note that the spontaneous generation  of a background field is  meant in the sense, that for the corresponding field the effective action has a minimum below zero, which is energetically favorable.

The paper is organized as follows. In next section \ref{T11}, we adduce  the representation of the effective potential from \cite{bord22-82-390} and note the main properties of it. In  section \ref{T4}, we investigate the minima of the effective potential in the pure magnetic case and high temperature.  In section \ref{T4a}, we consider the case of pure $A_0$ contribution and compare the energy of the $A_0$ and $b$ condensates accounting the one- and two-loop contributions for magnetic condensate. In section   \ref{T4b}, the conditions for stabilization of the charged gluon spectrum with accounting for both condensates are investigated. The last section \ref{T6} is devoted to discussion.

\ni Throughout the paper we use natural units with $\hbar=c=k_{\rm B}=1$.

\section{\label{T11}Representation for the effective potential}

In  the case of SU(2), the effective potential in the background $R_\xi$ gauge reads \cite{bord22-82-390}:
\eq{W2}{ W^{SU(2)}_{gl} &=B_4(0,0)+2B_4\left(a,b\right)
\\\nn&~~~	+2{g^2}\left[
	B_2\left(a,b\right)^2
	+2 B_2\left(0,b\right) B_2\left(a,b\right)	\right]
	-4{g^2}(1-\xi) B_3\left(a,b\right)B_1\left(a,b\right)
}
with the notation
\eq{ab}{a=\frac{x}{2}=\frac{g A_0}{2\pi T},~~~b=gH_3.
}
Since we work at finite temperature, $ W_{gl}$ is equivalent to the free energy.
The functions $B_n(a,b)$ are defined by
\eq{3}{ B_4(a,b) &= T\sum_\ell\int\frac{dk_3}{2\pi}\frac{b}{4\pi}\sum_{n,\sig}
	\ln\left(\left(2\pi T(\ell+a)\right)^2+k_3^2+b(2n+1+\sig-i0)\right),
\\\nn
B_3(a,b) &=
T\sum_\ell\int\frac{dk_3}{2\pi}\frac{b}{4\pi}\sum_{n,\sig}
\frac{\ell+a}{\left(2\pi T(\ell+a)\right)^2+k_3^2+b(2n+1+\sig-i0)}
\\\nn
	 B_2(a,b) &= T\sum_\ell\int\frac{dk_3}{2\pi}\frac{b}{4\pi}\sum_{n,\sig}
	\frac{1}{\left(2\pi T(\ell+a)\right)^2+k_3^2+b(2n+1+\sig-i0)},
\\\nn		
	B_1(a,b) &=
	T\sum_\ell\int\frac{dk_3}{2\pi}\frac{b}{4\pi}\sum_{n,\sig}
	\frac{\ell+a}{\left(\left(2\pi T(\ell+a)\right)^2+k_3^2+b(2n+1+\sig-i0)\right)^2}.
}
Here, $\xi$ is gauge fixing parameter,  the summations run  $n=0,1,\dots$, $\sig=\pm2$ and $\ell$ runs over all integers. The $'-i0'$-prescription  defines the sign of the imaginary part for the tachyon  mode.
These formulas and eq. \Ref{W2} are the generalization of the corresponding two-loop expressions in \cite{enqv90-47-291}, eqs. (3.8) and (A.2)-(A.5), \cite{bely91-254-153}, eq. (14),  \cite{skal92-7-2895}, eq. (4),  and also \cite{skal21-18-738}, eq. (4),
to the inclusion of the magnetic field. Note a "-" sign in \Ref{1.1}). Below we  use also the relations
\eq{3a}{B_3(a,b) &= \frac{1}{4\pi T}\pa_a B_4(a,b),
	~~~&	B_1(a,b) &= \frac{-1}{4\pi T}\pa_a B_2(a,b).
}

For $b\to 0$ we note $\frac{b}{4\pi}\sum_{n,\sig}\to\int\frac{d^2k}{(2\pi)^2}$ and get at $b=0$,
\eq{1.1}{ 	&&B_4(a,0)=\frac{2\pi^2 T^4}{3} B_4(a),
	 	~~	B_3(a,0)=\frac{2\pi T^3}{3}	B_3(a), \\ \nonumber
  		 &&B_2(a,0)=\frac{T^2}{2}	B_2(a),
	~~~~~~~	B_1(a,0)=-\frac{ T}{4\pi B_1(a)},
}
where $B_n(a)$ are the Bernoulli polynomials, periodically continued.
The special values for, in addition, $a=0$ are
\eq{4}{ B_4(0,0)  = -\frac{\pi^2T^4}{45},
	~~~   B_3(0,0)= 0,
	~~~   B_2(0,0)= \frac{T^2}{12},
	~~~   B_1(0,0)= \frac{T}{8\pi}.
}
We note that these formulas hold for $T>0$.
The motivation for the above choice of the notations  is that  the functions $B_n(a,b)$, \Ref{3}, are  the corresponding mode sums without additional factors. More details about this representation as well as the renormalization and the case of $T = 0$ are given in \cite{bord22-82-390}.



\section{\label{T4}The  magnetic field at high temperature}

Let us consider the case of high temperature. We use  eq.(59) of \cite{bord22-82-390}.
 The effective potential reads
\eq{3.6}{W^{SU(2)}_{gl}&=
	 \frac{b^2}{2 g^2}-\frac{\pi ^2 T^4}{15}
	 -\frac{a_1 b^{3/2} T}{2 \pi }
	 +\frac{11 b^2 \log (4 \pi  T/\mu)}{24 \pi ^2}
	 +g^2\left( \frac{T^4}{24}-\frac{{a_2} \sqrt{b} T^3}{12 \pi }
	 +\frac{{a_2}^2 b T^2}{32 \pi ^2}
	 \right)
}
The first term is the classical energy. The   terms proportional to $T^4$ constitute the gluon black body radiation. The contribution from the second loop is in the parenthesis. It has a $T^3$ behaviour. The numbers $a_1 = 0.828, a_2 = 1.856$ are calculated in \cite{bord22-82-390}, eq. (22).

In one-loop order, the energy \Ref{3.6} has a non-trivial minimum resulting from the term proportional to $b^{3/2}T$.   The condensate and the effective potential in its minimum are
\eq{3.7}{ b_{min}^{one} &= \frac{9 {a_1}^2 \alpha_s^2 T^2}{16 \pi ^2},
	&W^{SU(2), \,one}_{min}&=-\frac{\pi ^2 T^4}{15}	-\frac{27 a_1^4  \alpha_s^3 T^4}{512 \pi ^4},
}
where $\alpha_s = g^2/(1 + \frac{11}{12} \frac{g^2}{\pi^2} \log (4 \pi  T/\mu))$ is running coupling constant, $\mu $ is a normalization point for temperature.
The first term of the energy is the gluon black  body radiation. In this approximation, the condensate is  always  present, and the energy in the minimum is always negative. That means the spontaneous vacuum magnetization and SU(2) symmetry breaking. Here also an imaginary term presents, but we are concentrating on the real part. The standard way to remove the imaginary term of one-loop effective potential is adding the daisy diagram contributions (see \cite{skal00-576-430} for details). From \Ref{3.7} we see that the presence of $\alpha_s $ weakens the field strength at high temperature.

Now we turn to the two loop case. We consider the high temperature limit and take into consideration the $\sim T^3$ term in \Ref{3.6}. Denoting $b^{1/2} = x$ we obtain the third-order polynomial equation for determining the  condensate value:
\eq{3.8}{ x^3 - \frac{3}{4 \pi} a_1 T \alpha_s x^2 - \frac{g^2}{2 4 \pi} a_2 T^3 \alpha_s = 0. }
The real root of it can be found using formulas from the  standard handbook \cite{abra64}, Chapter 3.8. The result is
\eq{3.9}{ x_0 = b^{1/2}_{min} = \frac{1}{4}\frac{(2 a_2 \alpha_s)^{1/3}}{\pi^{1/3}} T + \frac{1}{4 \pi} a_1 \alpha_s T .}
If we compare this  with \Ref{3.7}, we find that the second term is three times less than the one in \Ref{3.7}. The most interesting is the change of the temperature dependence coming from $ \alpha_s^{1/3}$. Hence, the first term is dominant for this case.   For the field strength we get in this limit
\eq{3.10}{ b_{min} = \frac{1}{16}\frac{(2 a_2 \alpha_s)^{2/3}}{\pi^{2/3}} T^2.}
Note also, the value $a_2$ is larger than $a_1$. As a result, the role  of the second  loop is important.  As fare as we know, formula  \Ref{3.9} was not noted in the literature elsewhere.
%
\section{\label{T4a}The minimum of the effective potential for pure $A_0$ }
In this section, we remind the known results for the case of a pure $A_0$-background. We follow the recent paper \cite{bord21-81-998}, where this case was investigated in detail for $SU(3)$. In  the $SU(2)$,  formulas are simpler. For $b=0$, the effective action \Ref{W2} with \Ref{1.1} is expressed in terms of Bernoulli's polynomials. We restrict ourselves to the main topological sector and there to $0\le a\le 1/2$. Here, the effective potential has a minimum at $a= a_{min}$ (see also eq. (6) in \cite{skal92-7-2895}) and takes in this minimum the value $W_{|{a=a_{min}}}=W_{min}$ with
\eq{4.1}{(g A_0)_{min} &= \frac{3-\xi}{16\pi}g^2 T,
	& W_{min} &= -\frac{\pi^2T^4}{15}
	-\frac{(3-\xi)^2T^4}{192\pi^2}g^4.
}
As mentioned in \cite{skal21-18-738}, \cite{bord21-81-998}, eq. \Ref{4.1} coincides with the gauge-invariant result for $\xi=-1$, what we assume in the following. The first term of the effective potential is the gluon black body radiation.

Let us compare  \Ref{4.1} with the minimal effective potential \Ref{3.7} in the pure magnetic case. We see   in the latter case, the extra temperature dependent factor $(1 + \frac{11}{12} \frac{g^2}{\pi^2} [\log (4 \pi  T)/\mu)^{-1}$ is present and decreases the value of the magnetic condensate at high temperature. For the two loop result \Ref{3.10} the strength of the  field  is larger. But again at sufficiently high temperature the $ \alpha_s^{1/3}$ factor makes the value of $b_{min}(T)$ smaller compared to the value of $(g A_0)_{min}$ \Ref{4.1}.  As a result, since both condensates have negative energies they should  be generated. This decreases the total   free energy of the system.
%
%
\section{\label{T4b}Stability of the charged gluon spectrum}
The well known instability of gluon vacuum in magnetic field (which was not discussed above for special reasons) is the consequence of big magnetic moment for color charged gluons. For them the  gyromagnetic ratio equals to $\gamma = 2$. That results in the unstable (tachyon) mode   in the spectrum in magnetic field, $- p_4^2 = p_0^2 = p^2_3 + (2 n - 1) b$  for the lower state $n = 0$. In the presence of both condensates the charged gluon spectrum reads
\eq{4b.1}{(p_4 + (g A_0)_{min})^2 + p_3^2 + (2 n - 1) b_{min}.
}
First,  substituting  the  one loop minimum value \Ref{3.7} and \Ref{4.1} for $p_4 = 0, p_3 = 0 , n = 0$ we find the relation ensuring vacuum stability
\eq{4b.2}{ g^4   \geq  81 {a_1}^2 \alpha_s^2 .}
 Second, for the two loop case \Ref{3.10} we get
\eq{4b.3}{  g^4   \geq (2 a_2 \alpha_s)^{2/3} \pi^{4/3},}
 correspondingly. Clearly that the latter case is realized at more high temperature.  The vacuum consisting of two condensates is stable because of $ \alpha_s( T)$ decreasing with temperature increasing. This is the crucial fact coming from asymptotic freedom. Both condensates are needed to have stable vacuum. Because of different temperature dependence they dynamically  coexist in such a way that the effective potential is real in the considered approximation. There are no unstable modes at high temperature. Thus, the mechanism of vacuum stabilization is clarified.

\section{\label{T6}Conclusions}
We investigated  the two loop effective potential of $SU(2)$ gluodynamics in the background either $A_0$  or color magnetic field $H^3$. Using the integral representation \Ref{W2} - \Ref{3} we  calculated  the $A_0(T)$ and $H^3(T)$ condensates. For the former field, we have reproduced the known result from the literature (see resent paper \cite{skal21-18-738} (where the gauge invariance is also investigated) for more details). But for the latter one we  obtained new expression for the  field strength \Ref{3.9}. This result means  the importance of the two loop contribution to the effective potential and detects the temperature dependence of the magnetic field condensate.

It worth to note that the standard approximation for investigating  phase transitions at finite temperature accounting for the one loop plus daisy diagram contributions is insufficient in the case of two condensates. It is applicable  for the spontaneous magnetization, only. Remind,  the one-loop plus daisy diagrams is the consistent approximation in the order $g^2$ in coupling constant. The daisy diagrams account for long range correlations like tachyon states of the spectrum.  As it was found already, the imaginary part presenting   the one loop effective potential  is exactly cancelled by the term of the same order $g^2$ entering the daisy diagram series. As a result,  the effective potential is real in the given approximation (see, for instance, \cite{skal00-576-430}). But in such approximation there is no $A_0$ condensation.  The latter is realized as the correlation of the two- and one-loop contributions. Here it is important that $A_0(T)$ is dynamical parameter having zero value in tree approximation.

Within carried out investigations, we found the vacuum structure and the mechanism of the magnetic field stabilization at high temperature. It consists in the coexistence of the $A_0(T) $ and $H^3(T)$ condensates generated spontaneously. In this field configuration there are no unstable modes for charged gluons at high temperature.  This is in agreement with the heuristic speculations and qualitative estimates given in \cite{eber98-13-1723} and the results of the lattice simulations fulfilled already in \cite{demc08-41-165051}.

\end{document}